\documentclass[twocolumn,showpacs,preprintnumbers,amsmath,amssymb,floatfix,aps]{revtex4}

\newif\ifpdf
\ifx\pdfoutput\undefined
  \pdffalse 
\else
  \pdfoutput=1 
  \pdftrue
\fi

\ifpdf
  \usepackage[pdftex]{graphicx}
\else
  \usepackage[dvips]{graphicx}
  \usepackage{epsfig}
  \epsfclipon
\fi

\begin{document}
\title{The pion parton distribution function in the valence region}

\author{K. Wijesooriya}
\affiliation{Physics Division, Argonne National Laboratory, Argonne, IL  60439}

\author{P.E. Reimer}
\affiliation{Physics Division, Argonne National Laboratory, Argonne, IL  60439}

\author{R.J. Holt}
\affiliation{Physics Division, Argonne National Laboratory, Argonne, IL  60439}

\date{September 8, 2005}

\begin{abstract}
The parton distribution function of the pion in the valence region is
extracted in a next-to-leading order analysis from Fermilab E-615
pionic Drell-Yan data.  The effects of the parameterization of the
pion's valence distributions are examined.  Modern nucleon parton
distributions and nuclear corrections were used and possible effects
from higher twist contributions were considered in the analysis.  In
the next-to-leading order analysis, the high-$x$ dependence of the
pion structure function differs from that of the leading order
analysis, but not enough to agree with the expectations of pQCD and
Dyson-Schwinger calculations.
\end{abstract}

\maketitle

The pion has a central role in nucleon and nuclear structure.  The
pion has not only been used to explain the long-range nucleon-nucleon
interaction, but also to explain the flavor asymmetry observed in the
quark sea in the nucleon.  Experimental knowledge of the parton
structure of the pion arises primarily from pionic Drell-Yan
scattering~\cite{Conway:1989fs, Conway:1987hv, Badier:1983mj,
Betev:1985pf} from nucleons in heavy nuclei.  Recently, the shape of
the extracted pion parton distribution function (PDF) at high-$x$,
where $x$ is the fraction of the pion momentum carried by the
interacting quark, {\it i.e.}  Bjorken-$x$, has been
questioned~\cite{Hecht:2000xa}.

The anomalously light pion mass is believed to arise from dynamical
chiral symmetry breaking.  Any model of the pion must account for its
dual role as a quark-antiquark system and the Goldstone boson of
dynamical chiral symmetry breaking.  Theoretical descriptions of
pionic parton structure at high-$x$ disagree.  The parton
model~\cite{Farrar:1975yb}, perturbative quantum chromodynamics
(pQCD)~\cite{Ji:2004hz,Brodsky:1995kg} and some non-perturbative
calculations such as Dyson-Schwinger Equation (DSE)
models~\cite{Hecht:2000xa, Maris:2003vk, Bloch:1999ke, Bloch:1999rm,
Hecht:1999cr} indicate that the high-$x$ behavior should be near
$(1-x)^a$, with $a\gtrsim 2$.  In contrast, relativistic constituent
quark models~\cite{Frederico:1994dx, Szczepaniak:1994uq},
Nambu-Jona-Lasinio models with translationaly invariant
regularization~\cite{Shigetani:1993dx, Davidson:1995uv, Weigel:1999pc,
Bentz:1999gx}, Drell-Yan-West relation~\cite{Drell:1970km,
West:1970av} and even duality arguments~\cite{Melnitchouk:2002gh}
favor a linear high-$x$ dependence of $(1-x)$.  Instanton-based models
appear to lie in between these two~\cite{Dorokhov:2000gu}.  Lattice
calculations yield only the moments of the distributions and not the
PDFs themselves~\cite{Best:1997qp, Detmold:2003tm}.

While the PDFs for the nucleon are now well determined by global
analyses of a wide range of precise data (see {\it
e.g.}~\cite{Lai:1999wy, Martin:1998sq, Gluck:1998xa}) the pion PDFs
are very poorly known.  Presently there are only two experimental
techniques to access quark distributions in the pion: the deep
inelastic scattering from the virtual pion cloud of the proton for
which data is available in the low-$x$, sea region ($3 \times
10^{-4}\leq x \leq 0.01$) \cite{Adloff:1998yg, Chekanov:2002pf}, and
the Drell-Yan mechanism which provides data in the valence region
($0.2 \leq x \leq 0.99$).  Unfortunately there is no overlap between
these two experimental techniques.

A previous {\em leading order} (LO) analysis of pionic Drell-Yan data
by J. Conway {\it et al.} (Fermilab E615)~\cite{Conway:1989fs}
suggested a high-$x$ dependence of $(1-x)^{1.26}$.  The data were at
relatively high momentum transfer, $Q^2 > 16~\rm({GeV})^2$, where
pQCD should be a valid approach.  Previous analyses in both
LO~\cite{Conway:1989fs,Owens:1984zj} and next-to-leading order
(NLO)~\cite{Sutton:1992ay,Gluck:1999xe} have adopted a simple
functional form for the pionic valence quark distributions or made
other assumptions about the parton distributions.  Since the original
analysis of the Drell-Yan data, it has been
suggested~\cite{Hecht:2000xa} that the simple functional form assigned
to the valence quarks in the pion and that using only a LO analysis
might introduce a bias at high-$x$.

The purpose of this work is to re-analyze the pionic Drell-Yan cross
section data in order to study the high-$x$ behavior of the pion PDFs
and to see if such a bias exists. Cross section data from Fermilab
E615 (Conway {\it et al.} ~\cite{Conway:1989fs}) were fit to determine
the form of the high-$x$ pion parton distribution.  A NLO analysis
using the well understood PDFs for the proton
(MRST98~\cite{Martin:1998sq}, GRV98~\cite{Gluck:1998xa}, and
CTEQ5M~\cite{Lai:1999wy}, including nuclear
corrections~\cite{Eskola:1998df}) and allowing for higher twist
effects was performed.


The LO Drell-Yan cross section for a pion interacting
with a nucleon is
\begin{eqnarray}
\lefteqn{\frac{d^{2}\sigma}{dx_{\pi}dx_{\rm N}}  =  \frac{4\pi\alpha^2}{9M_\gamma^2} } \hspace{0.3in}\\ \nonumber
    &  \times  \sum e^2\left[q_\pi(x_\pi)\bar q_N(x_N) + \bar q_\pi(x_\pi) q_N(x_N)\right],
\end{eqnarray}
where the sum is over quark flavor, $q_{\pi{\rm (N)}}$ is the PDF for
quark flavor $q$ in the pion (nucleon); $e$ is the charge of the
quark, $M_\gamma$ is the mass of the virtual photon and
$x_{\pi{\rm(N)}}$ is the momentum fraction (Bjorken-$x$) of the
interacting quark in the pion (nucleon).  (Where not needed for
clarity, the subscript ``$\pi$'' will be dropped.)  The cross section
per nucleon on an atom with atomic number $Z$ and atomic mass $A$ is
\begin{equation}
\frac{d^{2}\sigma}{dx_{\pi}dx_{\rm N}} = \frac{Z}{A}\frac{d^{2}\sigma^{\rm prot.}}{dx_{\pi}dx_{\rm N}} + \left(1-\frac{Z}{A}\right)\frac{d^{2}\sigma^{\rm neut.}}{dx_{\pi}dx_{\rm N}}.
\end{equation}
Charge symmetry ({\it e.g.} $u_{\rm prot.} = d_{\rm neut.}$, etc.) was
then used to express the cross section in terms of just the pion and
proton's PDFs.

The pion's valence ($q_\pi^{\rm v}$), sea($q_\pi^{\rm s}$) and
gluonic($g_\pi$) parton distributions were parameterized as
\begin{eqnarray}
x q_\pi^{\rm v}(x) & = & A_\pi^{\rm v} 
                                 \left[x^\alpha
                                      \left(1-x\right)^\beta
                                      \left(1-\epsilon\sqrt{x}+\nu x\right)\vphantom{\frac{2x^2}{9m^2_{\mu\mu}}}\right.\\\nonumber
& & \left.\hphantom{A_\pi^{\rm v} \left[\right.}
                                      + \gamma\frac{2x^2}{9m^2_{\gamma\gamma}}\right],\\
x q_\pi^{\rm s}(x) & = & A_\pi^{\rm s} \left(1-x\right)^\delta{\rm ~and}\\
x g_\pi(x) & = & A^{\rm g}_\pi \left(1-x\right)^\eta,
\end{eqnarray}
respectively.  The valence parameterization, $x q_\pi^{\rm v}$,
follows that suggested by Hecht {\it et al.}~\cite{Hecht:2000xa} with
the addition of the term proportional to $x^2/m^{2}_{\mu \mu}$ that
allows for the possibility of higher-twist effects as suggested by
Berger and Brodsky~\cite{Berger:1979du} and used by Conway {\it et
al.}~\cite{Conway:1987hv, Conway:1989fs}.  The coefficient of this
term, $\gamma$, was restricted to only positive values.  This
parameterization reduces to the ``minimal parameterization'' used in
earlier works~\cite{Conway:1987hv, Conway:1989fs} if $\epsilon$ and
$\nu$ are both fixed at 0.  If $\beta = 1$, the pion valence PDF has a
linear high-$x$ behavior and as $\beta$ increases, there is more
curvature at high-$x$.  The pion's sea and gluon distributions are
identical to those used in earlier works.  We assume the pion's
valence and sea distributions are SU(3) flavor symmetric:
\begin{eqnarray}
x q_\pi^{\rm v}(x) & = & x \bar{u}_\pi^{\rm v}(x)  =  x d_\pi^{\rm v}(x) {\rm ~and} \\
\label{eqn:su3}
x q_\pi^{\rm s}(x) & = & x \bar{u}_\pi^{\rm s}(x)  =  x u_\pi^{\rm s}(x)\\ \nonumber 
                           & = & x \bar{d}_\pi^{\rm s}(x)  =  x d_\pi^{\rm s}(x) \\ \nonumber
                           & = & x \bar{s}_\pi^{\rm s}(x)  =  x s_\pi^{\rm s}(x).
\end{eqnarray}
The sea and gluonic distributions are better determined from other
data.  The shape of the gluon distribution is deduced from CERN WA70
prompt photon data~\cite{Bonesini:1987mq}.  A fit to these data is
it best described by $\eta = 2.1$ and $G_\pi \equiv \int_0^1 x
g_\pi(x) dx = 0.47$~\cite{Sutton:1992ay}.  The sea
distributions can be determined by comparing $\pi^+$ and $\pi^-$
induced Drell-Yan scattering as was done by the CERN NA3 collaboration
who found $\delta = 8.4$~\cite{Badier:1983mj}.  These parameters were
fixed at the values given above and were also used by Conway {\it
et al.}~\cite{Conway:1989fs}.

Sum rules constrain the normalization coefficients $A^{\rm v}_{\pi}$
and $A^{\rm s}_{\pi}$.  Specifically, the total number of valence
$\bar{u}$ (and $d$) quarks in the $\pi^-$ is constrained to be unity
by
\begin{equation}
\int_{0}^{1} q^{\rm v}_{\pi}\left(x_{\pi}\right) dx_{\pi} = 1.
\label{eq:constraint}
\end{equation}
Momentum conservation within the pion is enforced by requiring
\begin{equation}
2 \int_{0}^{1} x q^{\rm v}_{\pi}\left(x_{\pi}\right) dx_{\pi} +6 \int_{0}^{1}\
 x q^{\rm s}_{\pi}\left(x_{\pi}\right) dx_{\pi} + G_{\pi} = 1.
\end{equation}

\begin{figure}

  \begin{center}
    \includegraphics[width=\columnwidth]{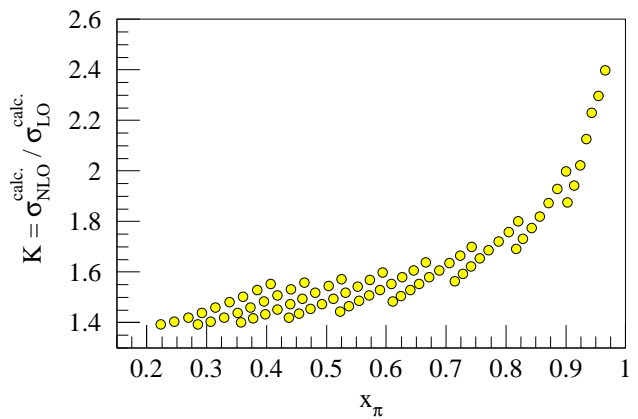}
  \end{center}

  \caption{(Color online) The ratio of NLO to LO pionic Drell-Yan
  cross sections ($K$-factor) calculated using CTEQ5M proton PDF and
  the pion PDF from this work. One point is shown for each
  $(x_F,\sqrt{\tau})$ bin projected onto the $x_\pi$ axis.  NLO terms
  are clearly important as $x\rightarrow 1$.}

  \label{fig:kfacth}

\end{figure}

Previous experiments have shown that the basic, LO Drell-Yan
cross-section formula fails to explain the magnitude of the observed
cross-section, $\sigma^{\rm exp}$, by a factor of nearly two.  This is
traditionally accounted for by multiplying the LO cross section,
$\sigma^{\rm LO}$ by a ``$K$''-factor, {\it i.e.}  $\sigma^{\rm exp} =
K\sigma^{\rm LO}$, which is completely independent of kinematics.
This difference is attributed to NLO terms and in proton-nucleon
Drell-Yan, $K^\prime \approx 1$, where $\sigma^{\rm exp} = K^\prime
\sigma^{\rm NLO}$~\cite{Webb:2003ps}.  Comparisons of LO and NLO cross
section calculations show that the ``$K$''-factor is not completely
independent of kinematics, in particular in the high-$x$ region,
as shown in Fig.~\ref{fig:kfacth}.  Note that $K$ and $K^\prime$ are
indistinguishable from normalization errors.  In the NLO fits, it was
assumed that the NLO calculation accounted for the entire observed
cross section, and so $K^\prime$ reflected the normalization
uncertainty.  Thus, deviation of $K^\prime$ from one contributed to
the $\chi^2$ of the NLO fits, while $K$ was allowed to vary freely in
the LO fits.

The differential cross-section measured by Conway {\it et
al.}~\cite{Conway:1989fs} were the basis for this work.  Events
arising from $J/\psi$ and $\Upsilon$ decays were removed by requiring
that $4.05 \leq M_\gamma \leq 8.53~\textrm{GeV}$.  (Relaxing this
condition to include bins with $M_\gamma \geq 11.06~\textrm{GeV}$
had no noticeable effect on the results.)  Only bins with $x_F \geq 0$
(Feynman-$x$) were included in the fit.  Furthermore to minimize the
effects of shadowing in copper, only bins in which $x_{\rm N} \geq
0.06$ were used.  Finally, if the center of the bin was not within the
defined acceptance that bin was not considered by the fit.  It is
important to note that this final restriction only removed six bins;
however, three of these had $x_\pi\ge0.97$.  These same restrictions
were used in the original fit of these data~\cite{Conway:1987hv,
Conway:1989fs}.  With these restrictions, data from 76 bins of the 168
tabulated in Ref.~\cite{Conway:1989fs} remained.  The averaged mass
for the data included in the fit was 5.2 GeV.
\begin{table*}

  \caption{Tabulated below are the results from the leading order (LO)
  and next-to-leading order (NLO) fits at a mass scale of $\langle
  M_\gamma \rangle^2 = (5.2~\textrm{GeV})^2$.  The original LO fit
  of Conway {\it et al.}~\cite{Conway:1989fs} is tabulated in the
  first column.  The second column is a LO fit using the Conway {\it
  et al.}~\cite{Conway:1989fs} parameterization of the proton.
  Succeeding three columns are based on the CTEQ5L parameterization of
  the proton showing the effects of nuclear corrections and the
  extended pion PDF parameterization.  Following the LO results are
  the results of the NLO fits.  The sixth column gives the results for
  the full parameterization while the seventh column shows the
  ``minimal'' parameterization results (preferred fit).  The next two
  columns give the results for the ``minimal'' parameterization using
  the MRST and GRV proton parameterizations.  In the final column we
  see the effect of eliminating the higher twist contributions.  In
  all cases, $\delta$ and $g_\pi$ were fixed at $\delta = 8.40$ and
  $g_\pi = 0.47$.  \label{tab:fit_results}}

 \begin{tabular}{c|
  r@{$\pm$}l 
  r@{$\pm$}l 
  r@{$\pm$}l 
  r@{$\pm$}l 
  r@{$\pm$}l |
  r@{$\pm$}l 
  r@{$\pm$}l 
  r@{$\pm$}l 
  r@{$\pm$}l 
  r@{$\pm$}l 
 }
\hline \hline
 & \multicolumn{10}{|c}{Leading Order Fit Results}
 & \multicolumn{10}{|c}{Next-to-Leading Order Fit Results} \\
Prot. PDF 
 & \multicolumn{2}{c}{Fit from} 
 & \multicolumn{2}{c}{Conway} 
 & \multicolumn{2}{c}{CTEQ5L}
 & \multicolumn{2}{c}{CTEQ5L}
 & \multicolumn{2}{c|}{CTEQ5L} 
 & \multicolumn{2}{c}{CTEQ5M} 
 & \multicolumn{2}{c}{CTEQ5M} 
 & \multicolumn{2}{c}{MRST98}
 & \multicolumn{2}{c}{GRV98} 
 & \multicolumn{2}{c}{CTEQ5M}
\\
Nucl. Corr.
 & \multicolumn{2}{c}{Conway~\cite{Conway:1989fs}}
 & \multicolumn{2}{c}{-}
 & \multicolumn{2}{c}{-}
 & \multicolumn{2}{c}{EKS98}
 & \multicolumn{2}{c|}{EKS98} 
 & \multicolumn{2}{c}{EKS98}
 & \multicolumn{2}{c}{EKS98}
 & \multicolumn{2}{c}{EKS98}
 & \multicolumn{2}{c}{EKS98}
 & \multicolumn{2}{c}{EKS98} 
\\ \hline
$\alpha$ 
 &    0.60 &    0.03
 &    0.65 &    0.07
 &    0.67 &    0.07
 &    0.67 &    0.07
 &    0.43 &    0.17
 &    0.43 &    0.05
 &    0.70 &    0.06
 &    0.69 &    0.06
 &    0.69 &    0.06
 &    0.66 &    0.04
\\
$\beta$ 
 &    1.26 &    0.04
 &    1.33 &    0.08
 &    1.35 &    0.08
 &    1.36 &    0.08
 &    1.38 &    0.26
 &    1.60 &    0.08
 &    1.54 &    0.08
 &    1.54 &    0.08
 &    1.53 &    0.08
 &    1.46 &    0.04
\\
$\epsilon$ 
 & \multicolumn{2}{c}{\it    0.00}
 & \multicolumn{2}{c}{\it    0.00}
 & \multicolumn{2}{c}{\it    0.00}
 & \multicolumn{2}{c}{\it    0.00}
 &   -3.39 &    7.02
 &   -3.30 &    1.90
 & \multicolumn{2}{c}{\it    0.00}
 & \multicolumn{2}{c}{\it    0.00}
 & \multicolumn{2}{c}{\it    0.00}
 & \multicolumn{2}{c}{\it    0.00}
\\
$\nu$ 
 & \multicolumn{2}{c}{\it    0.00}
 & \multicolumn{2}{c}{\it    0.00}
 & \multicolumn{2}{c}{\it    0.00}
 & \multicolumn{2}{c}{\it    0.00}
 &   -0.76 &    5.52
 &   -0.01 &    1.35
 & \multicolumn{2}{c}{\it    0.00}
 & \multicolumn{2}{c}{\it    0.00}
 & \multicolumn{2}{c}{\it    0.00}
 & \multicolumn{2}{c}{\it    0.00}
\\
$\gamma$ 
 &    0.83 &    0.26
 &    0.78 &    0.65
 &    0.75 &    0.63
 &    0.77 &    0.62
 &    3.0  &    2.8\footnote{Converged at upper limit of parameter.}
 &    3.0  &    2.0$^a$
 &    0.60 &    0.34
 &    0.57 &    0.34
 &    0.57 &    0.34
 & \multicolumn{2}{c}{\it    0.00}
\\
$K$ ($K^\prime$) 
 &    1.75 &    0.13
 &    1.57 &    0.11
 &    1.53 &    0.10
 &    1.49 &    0.10
 &    1.56 &    0.24
 &    1.02 &    0.05
 &    0.97 &    0.06
 &    0.98 &    0.06
 &    0.98 &    0.06
 &    1.01 &    0.05
\\
$\chi^2/N_{DF}$ 
 & \multicolumn{2}{c}{ 359/329}
 & \multicolumn{2}{c}{  67.6/72}
 & \multicolumn{2}{c}{  67.1/72}
 & \multicolumn{2}{c}{  69.3/72}
 & \multicolumn{2}{c|}{  69.1/70}
 & \multicolumn{2}{c}{  72.8/70}
 & \multicolumn{2}{c}{  73.1/72}
 & \multicolumn{2}{c}{  72.0/72}
 & \multicolumn{2}{c}{  70.9/72}
 & \multicolumn{2}{c}{  74.4/73}
\\ \hline \hline
 \end{tabular}
\end{table*}

The data were fit in both NLO and LO.  In both cases, no QCD evolution
was performed so that the resulting pion PDFs are at the
average mass scale of the data, $\langle M_\gamma
\rangle^2 = (5.2~\textrm{GeV})^2$.  The LO fits were used to
compare with the fit of Conway {\it et al.}~\cite{Conway:1989fs} and
so that the effects of the NLO terms in the cross section could be
clearly understood.  The results shown in Tab.~\ref{tab:fit_results}.
All of the LO fits have slightly more curvature at high-$x$ than the
original fit of these data~\cite{Conway:1989fs}.  This includes the
one that used the Conway parameterization of the proton.  The original
work by Conway {\it et al.} was based on a finer binning of the data,
but the largest differences between the two fits are in the
$K$-factor.  In this context, it should be noted that a direct
calculation based on Conway {\it et al.}'s parameterizations of the
proton and pion reveal that $K=\sigma_{NLO}/\sigma_{LO} \approx 1.5$
in close agreement with the present fit and a kinematic dependence
similar to that in Fig.~\ref{fig:kfacth}. Fixing $K = 1.75$, more
closely reproduces Conway {\it et al.}'s fit with $\alpha = 0.56\pm
0.01$ and $\beta = 1.24\pm 0.05$.  The use of modern parton
distributions and the inclusion of nuclear corrections have only a
slight effect on the high-$x$ PDF.  In LO when
$\epsilon$ and $\nu$ were allowed to freely vary there were
significant correlations between fit parameters with $\epsilon$
becoming large and negative while $\gamma$ became large and positive
for trivial gains in $\chi^2$.  A bound of $0 < \gamma < 3.0$ was
implemented in the fit.  The shallowness of the $\chi^2$ hyper-surface
is evident in the large uncertainties in these parameters.

\begin{figure*}

  \begin{center}
    \includegraphics[width=\textwidth]{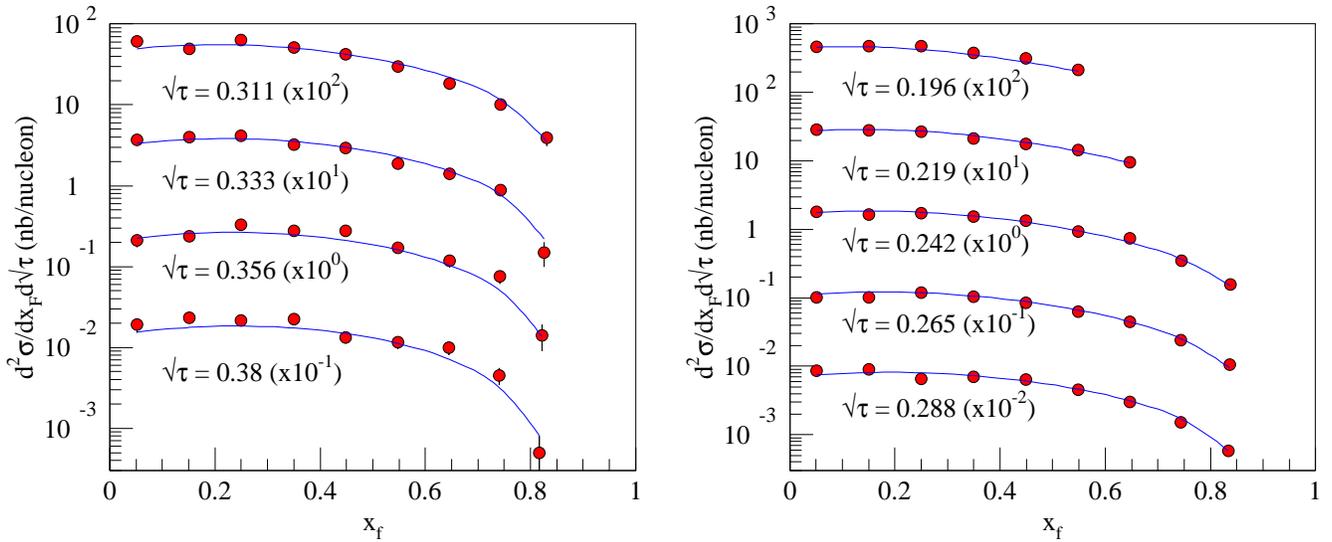}
  \end{center}

  \caption{(Color online) Measured values of
  $d\sigma/dx_Fd\sqrt{\tau}$ (points) reported by Conway {\it et
  al.}~\cite{Conway:1989fs} along with the NLO fit to the data
  (curves).}

  \label{fig:sigma}

\end{figure*}

\begin{figure}

  \begin{center}
    \includegraphics[width=\columnwidth]{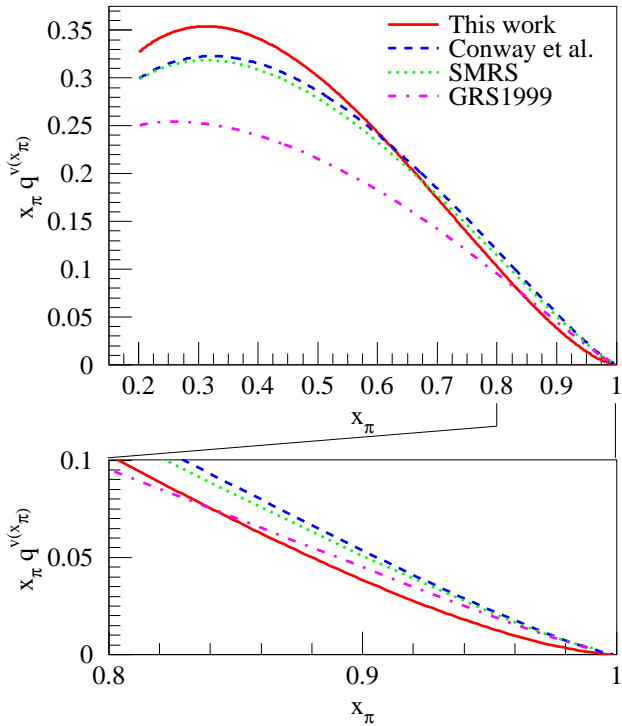}
  \end{center}

  \caption{(Color online) The valence pion PDF, $x q^v_\pi(x)$, for
  the preferred fit (solid) as well as those from Conway, {\it et
  al.}~\cite{Conway:1989fs} (dashed), SMRS~\cite{Sutton:1992ay}
  (dotted) and GRS~\cite{Gluck:1998xa} (dot-dashed) at a mass scale of
  $\langle M_\gamma \rangle^2 = (5.2~\textrm{GeV})^2$ are
  shown. The present work shows evidence for greater curvature at
  high-$x$.  The upper plot shows the entire $x$ range of the data,
  while the lower plot emphasized the high-$x$ region.  For
  comparison, the small higher twist ($\gamma$) term has been
  removed.}

  \label{fig:fitresults}

\end{figure}

These fits were repeated in NLO using the CTEQ5M proton
parameterization.  As expected from the kinematic dependence of the
$K$-factor (see Fig.~\ref{fig:kfacth}) there is even more curvature at
high-$x$ in both the ``minimal'' and full parameterization with $\beta
= 1.54$ and $\beta = 1.60$ respectively.  As in the LO case, the fit
to the full parameterization was able to offset increases in $\gamma$
(the higher twist term) with changes in $\epsilon$ and $\nu$ for {\it
extremely} slight gains in $\chi^2$.  Based on this, and the
relatively large uncertainties in these three parameters, it was
reasonable to remove either $\gamma$ from the fit or fix $\epsilon=0$
and $\nu=0$.  Fixing $\gamma$ at values between $0.0$ and $0.8$ yields
fits which have {\it significantly} larger uncertainties in $\epsilon$
and $\nu$.  Alternatively, $\gamma$ has a direct physical
interpretation, while $\epsilon$ and $\nu$ dilute the interpretation
of $\beta$ and are merely present to allow for a better representation
of the data.  While not considered in this work, the angular
distributions observed by Conway {\it et al.}  and by S. Falciano {\it
et al.}  (CERN NA10)~\cite{Falciano:1986wk} are better described with
the inclusion of the higher twist term.  Using the ``minimal''
parameterization (fixing $\epsilon = 0$ and $\nu = 0$) and allowing
$\gamma$ to vary fits the data as well as allowing all three
parameters to vary.  In fact, the distributions of residuals as a
function of $x$ are nearly identical to the full fit.  Based on these
considerations, the ``minimal'' parameterization was adopted as the
{\it preferred} parameterization.  This fit, compared to data is shown
in Fig.~\ref{fig:sigma} and the resulting pion valence parton
distribution in Fig.~\ref{fig:fitresults}.  To facilitate the
comparison of this fit with lattice results, Tab~\ref{tab:moments}
gives the moments, $\int^1_0 x^n q^{\rm v}_\pi(x) dx$, of the pion's
valence distributions.

\begin{figure}

  \begin{center}
    \includegraphics{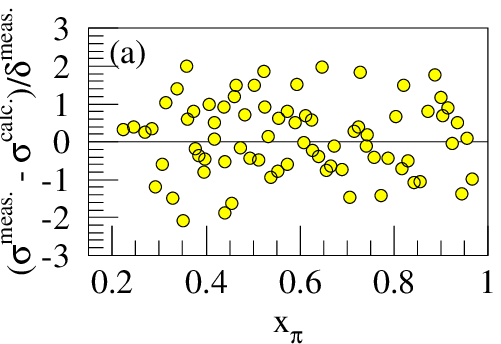}
    \includegraphics{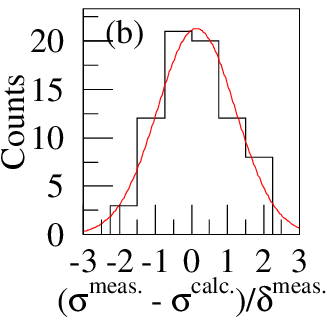}
  \end{center}

  \caption{(Color online) (a) The residuals, $R = \left(\sigma^{\rm
  meas.}  - \sigma^{\rm calc.}\right)/\delta^{\rm meas.}$, for the
  preferred fit as a function of $x$.  No bias is apparent as a
  function of $x$.  (b) The residual distribution fit to a Gaussian
  with $\langle R \rangle = 0.14 \pm 0.15$ and $\sigma = 1.09 \pm
  0.14$.}

  \label{fig:residuals}

\end{figure}

To determine if there was any bias in the fit, it is useful to examine
the residual or pull of each data point defined by $R =
\left(\sigma^{\rm meas.} - \sigma^{\rm calc.}\right)/\delta^{\rm
meas.}$ where $\sigma^{\rm meas. (calc.)}$ is the measured
(calculated) cross section and $\delta^{\rm data}$ is the uncertainty
of the measured cross section.  The distribution of the residuals as a
function of $x$ reveals no bias, shown in Fig.~\ref{fig:residuals}.
The overall distribution is well described by a Gaussian distribution
with a standard deviation consistent with unity and mean with zero as
shown in Fig.~\ref{fig:residuals}.

\begin{table}

  \caption{The moments of the pion valence PDF, evaluated at a mass
  scale of $\langle M_\gamma \rangle^2 = (5.2~\textrm{GeV})^2$,
  are tabulated below.  The $0^{\rm th}$ moment was a constraint of
  the fit--see Eq.~\ref{eq:constraint}.\label{tab:moments} }

  \begin{tabular}{cr@{$\pm$}l}
\hline \hline
n & \multicolumn{2}{c}{\hspace*{0.05in}$\displaystyle\int^1_0 x^n q^{\rm v}_\pi(x) dx$ } \\ \hline
1 & 0.217  & 0.011\\
2 & 0.087  & 0.005 \\
3 & 0.045  & 0.003 \\
\hline \hline
  \end{tabular}

\end{table}
For comparison of the influence of the proton PDF on the fit,
additional NLO fits were performed using the MRST~\cite{Martin:1998sq}
and GRV98~\cite{Gluck:1998xa} parameterizations.  These fits produced
results similar to those obtained using the CTEQ parameterization, as
shown in Tab.~\ref{tab:fit_results}.

The contribution of the pion's strange sea was also investigated.  The
strength of the pion's strange sea was varied from being equal to the
up and down sea [SU(3) flavor symmetric limit--see Eq.~\ref{eqn:su3}]
to no strange sea.  Since most of the data was in the pion's valence
region, it is not surprising that the strange sea of the pion had
little effect on the valence parameterization.

Several earlier works have also considered the data fit by the present
work, as shown in Fig.~\ref{fig:fitresults}.  The fit by Conway {\it
et al.} was in LO and showed less curvature at high-$x$, but Conway
even speculated that a NLO fit might yield different
results~\cite{Conway:1987hv}.  The fit by SMRS~\cite{Sutton:1992ay}
looked at this data as well as other pionic data.  While this fit was
in NLO, it was based on what is now an old description of the proton's
PDFs with a linear, {\it ad hoc} nuclear correction.  In addition,
some of the high-$\tau$ data was excluded from the fit.
GRS~\cite{Gluck:1998xa} studied the pionic PDF in NLO
with the less restrictive parameterization, but this work also invoked
``constituent quark independence'' which directly relates the pion and
proton distributions~\cite{Gluck:1998xa}, introducing a different
possible bias.

We have presented a NLO analysis of Drell-Yan data from Fermilab
E-615.  When analyzed in NLO, the pions valence PDF
clearly have more curvature as $x \rightarrow 1$ than previous leading
order fits.  This is in worse agreement than previous analysis with
the linear dependence expected by NJL models~\cite{Shigetani:1993dx,
Davidson:1995uv, Weigel:1999pc, Bentz:1999gx}, duality
argument~\cite{Melnitchouk:2002gh} and the Drell-Yan-West
relation~\cite{Drell:1970km, West:1970av}.  It is not as much as is
expected from Dyson-Schwinger Equation based models of the
pion~\cite{Hecht:2000xa,Maris:2003vk,Bloch:1999ke,Bloch:1999rm,Hecht:1999cr}
or pQCD~\cite{Ji:2004hz,Brodsky:1995kg}, however.  Finally, while
E615's resolution in $x$ is not given, the mass resolution for the
$J/\psi$ is $\sigma_M = 0.16~\textrm{GeV}$~\cite{Conway:1987hv}.
This could translate into substantial uncertainty in $x$ as
$x\rightarrow 1$, so additional data from a more precise experiment
with better resolution would certainly be welcome.

We thank C.D.~Roberts and W.~Melnitchouk for many useful discussions
and W. Tung of CTEQ for providing the code necessary for the NLO
Drell-Yan cross section calculation.  This work was supported by the
U.S.\ Department of Energy, Office of Nuclear Physics, under Contract
No. W-31-109-ENG-38.

\bibliography{dypion}

\end{document}